\documentclass[aps,pre,twocolumn,superscriptaddress,showkeys,showpacs]{revtex4-1}
\usepackage{graphicx}% Include figure files
\usepackage{dcolumn}% Align table columns on decimal point
\usepackage{bm}% bold math
\usepackage{xcolor}
\usepackage{lineno}
\usepackage[capitalize]{cleveref}

\usepackage{lineno}

\begin{document}

%\preprint{APS/123-QED}

\title{Coupled dynamical phase transitions in driven disk packings}% Force line breaks with \\

\author{Akash Ghosh}
\affiliation {Dept. of Condensed Matter Physics and Materials Science, Tata Institute of Fundamental Research, Mumbai 400005, India}
\author{Jaikumar Radhakrishnan}
\affiliation{School of Technology and Computer Science,  Tata Institute of Fundamental Research, Mumbai 400 005, India}
\author{Paul M. Chaikin}
\affiliation {Center for Soft Matter Research and Department of Physics, New York University, New York, NY 10003}

\author{Dov Levine}
\affiliation { Department of Physics, Technion-IIT,  Haifa 32000, Israel}
\author{Shankar Ghosh  }
%\affiliation {Department of Condensed Matter Physics and Materials Science, Tata Institute of Fundamental Research, Mumbai 400005, India}
\affiliation {Dept. of Condensed Matter Physics and Materials Science, Tata Institute of Fundamental Research, Mumbai 400005, India}

% \email{sghosh@tifr.res.in}
\date{\today}% It is always \today, today,
%\linenumbers
\begin{abstract}

Under the influence of oscillatory shear, a  mono-layer of frictional granular disks exhibits two dynamical phase transitions: a transition from an initially disordered state to an ordered crystalline state, and a dynamic active-absorbing phase transition. Although there is no reason, {\it a-priori}, for these to be at the same critical point, they are. The transitions may also be characterized by the disk trajectories, which are non-trivial loops breaking time-reversal invariance.
%\begin{description}
%\end{description}
\end{abstract}

\maketitle

%\tableofcontents

%\section{\label{sec:level1} Introduction}
%\linenumbers

Dynamical phase transitions between active and absorbing states  have been  a paradigm in non-equilibrium physics for some time~\cite{Non-Equilibrium_Book,hinrichsen_non-equilibrium_2000}. 
 %One of the modern archetypes of non-equilibrium physics is a dynamical phase transition between active and absorbing states ~\cite{Non-Equilibrium_Book,hinrichsen_non-equilibrium_2000}. 
On one side of the transition the system is dynamic, where a fraction $\alpha$ of the particles are ``active", and move continually.  On the other side, all motion ceases.  Typically, this cessation of dynamics occurs when the constituent particles achieve some local goal, such as  becoming non-overlapping, or having no near neighbors.  The transition is characterized by a critical point at which long-range correlations are manifested, but the critical state is typically disordered and hyperuniform~\cite{torquato_local_2003}, as has been shown~\cite{PRL_hyper} for systems in the conserved directed percolation (or Manna) class~\cite{Non-Equilibrium_Book}. Such a transition has been studied in a variety of models theoretically~\cite{Non-Equilibrium_Book}, and experimentally in a system of non-Brownian hard spheres suspended in a viscous fluid~\cite{pine_chaos_2005,corte2008random,  wilken2020hyperuniform}.  For these systems,  the  fraction $\alpha$ of active particles  is the order parameter for the transition:  in the absorbing phase, $\alpha = 0$; in the active phase $\alpha > 0$.

% One of the modern paradigms of non-equilibrium physics is a dynamical phase transition between active and absorbing states.\cite{Non-Equilibrium_Book,hinrichsen_non-equilibrium_2000}. On one side of the transition the system is dynamic, where a fraction $\alpha$ of the particles are ``active", and move continually.  On the other side, all motion ceases.  Typically, this cessation of dynamics occurs when the constituent particles achieve some local goal, such as not overlapping, or having no near neighbors.  The transition is characterized by a critical point at which long-range correlations are manifested, but the critical state is typically disordered and hyperuniform\cite{torquato_local_2003}, as has been shown\cite{PRL_hyper} for systems in the conserved directed percolation (or Manna) class\cite{Non-Equilibrium_Book}. Such a transition has been studied in a variety of models theoretically\cite{Non-Equilibrium_Book}, and experimentally in a system of non-Brownian hard spheres suspended in a viscous fluid\cite{pine_chaos_2005,corte2008random}.  For these systems, the order parameter for the transition is $\alpha$:  In the absorbing phase, $\alpha = 0$; in the active phase $\alpha > 0$. $\alpha$ is typically called the active fraction, or the activity.

Many  of these investigations involve   studying  disordered systems  under 
 cyclic shear~\cite{ren2013reynolds,regev2015reversibility,kawasaki2016macroscopic,bhaumik2021role,leishangthem2017yielding,parmar2019strain, fiocco2014encoding,mukherji2019strength,paulsen2014multiple,ness2020absorbing,mangan2008reversible,reichhardt2022reversible}, where  the absorbing phase  is linked to the system's elastic response and the commencement of the active phase is related to yielding phenomena ~\cite{kawasaki2016macroscopic,bhaumik2021role,leishangthem2017yielding,parmar2019strain,yeh2020glass,mungan2021metastability}.  In this phase the constituent particles  can exhibit interesting limit cycle dynamics~\cite{ren2013reynolds,regev2015reversibility,regev2013onset}.   Such continued  cyclic  shearing   can modify the  yielding transition~\cite{kawasaki2016macroscopic,royer2015precisely} and even encode the  memory of the training  process~\cite{ren2013reynolds, fiocco2014encoding,mukherji2019strength,paulsen2014multiple}. 

% \textcolor{blue}{ In several of these investigations, cyclic shear in disordered system is used \cite{ren2013reynolds,regev2015reversibility,kawasaki2016macroscopic,bhaumik2021role,leishangthem2017yielding,parmar2019strain, fiocco2014encoding,mukherji2019strength,paulsen2014multiple,ness2020absorbing,mangan2008reversible}. The absorbing phase  is linked to the system's elastic response.  In this phase the constituent particles  can exhibit limit cycle dynamics \cite{ren2013reynolds,regev2015reversibility,regev2013onset}. The commencement of the non-absorbing phase is related to the  yielding phenomena \cite{kawasaki2016macroscopic,bhaumik2021role,leishangthem2017yielding,parmar2019strain,yeh2020glass,mungan2021metastability}. The system anneals as a result of the continual cyclic shearing, and the yielding transition which can modified by the shearing protocol is achieved by increasing the shearing amplitude. This transition is  accompanied by a  sudden change in the single particle  mobility \cite{kawasaki2016macroscopic,royer2015precisely}. Furthermore, investigations in athermal situations reveal that the memory of the shearing amplitude  can be encoded into the system by training it  \cite{ren2013reynolds, fiocco2014encoding,mukherji2019strength,paulsen2014multiple}. In this context, the absorbing to non-absorbing transition  is linked to the underlying jamming transition \cite{ness2020absorbing}.
% }

\begin{figure}[t]
	\centering
	\includegraphics[width=1\linewidth]{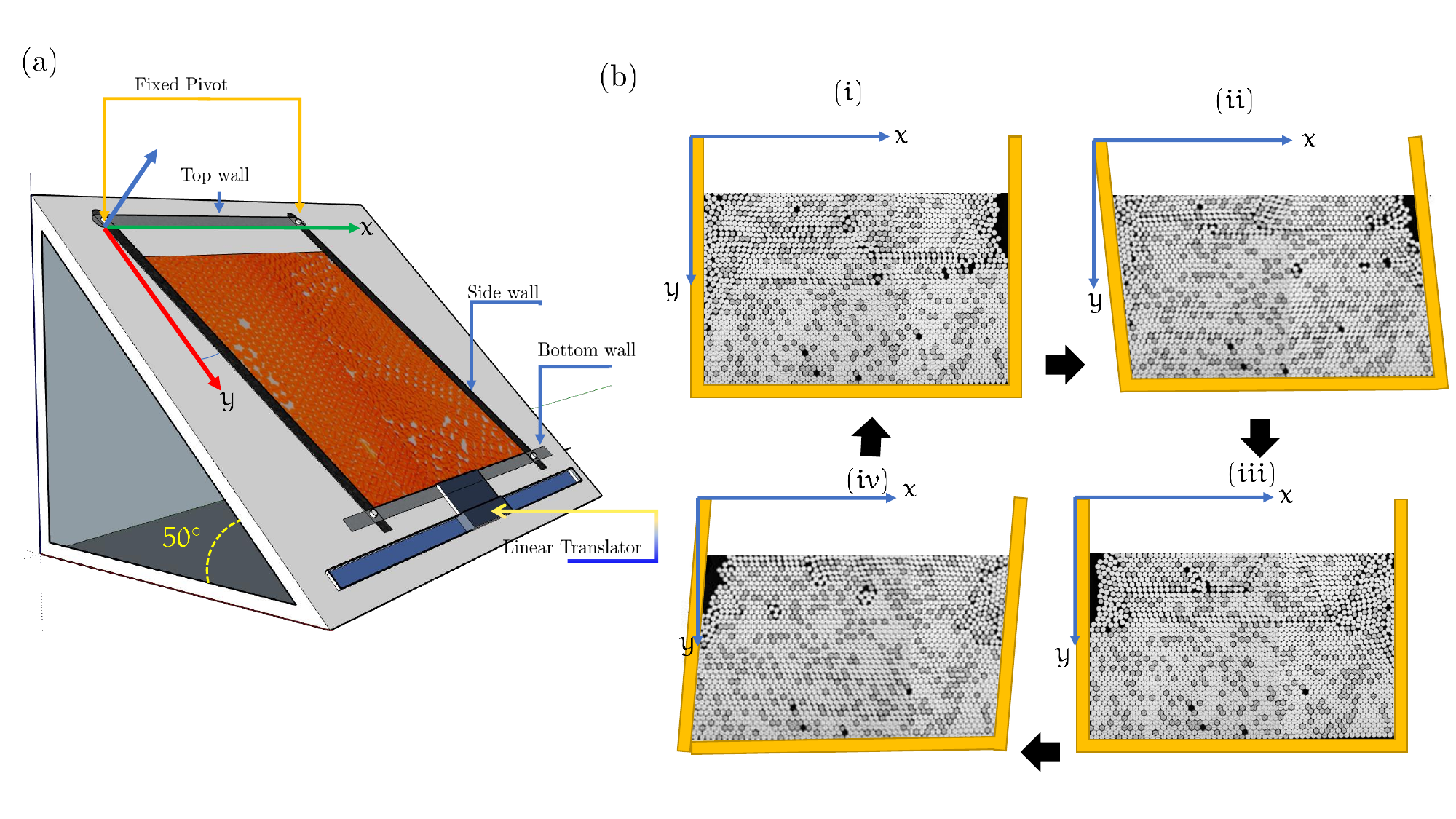}
	\caption{(a) Schematic of the apparatus.  All the walls of the shear cell are about 53 particle diameters long. (b) One cycle: The sequence $(i)\rightarrow(ii)\rightarrow(iii)\rightarrow(iv)\rightarrow (iv) \rightarrow (i) $  completes one shear cycle.  The sequence  starts from the ``zero position"  $(i)$, moving to the rightmost position ($ ii $), returning   to the central position  ($ iii $), then moving on to the leftmost position ($ iv $) before finally coming back to the zero position.  }
	\label{fig:Fig1}
\end{figure}

In this paper, we report   our findings  from   experiments performed with a collection of frictional disks that are subjected to oscillatory shear.  The system, depicted in Fig.~\ref{fig:Fig1}, consists of a single layer of identical  disks  confined between glass plates which are inclined with respect to the normal.  The disks interact frictionally with one another as well as with the bottom plate. In addition, they experience gravity, as a result of which there is a pressure gradient in the system \footnote{We note that this is true in a coarse-grained sense. On a grain scale, we expect force chains, although because of the continual shearing, these should be rather short in extent and in temporal duration.}. As we will detail presently, for large strain the system may be regarded as being in an active state, where a fraction of the disks  do not return to their positions after a strain cycle. For small strain, the system goes to an absorbing state, where all the disks return to their initial positions.  The system, however, is typically not homogeneous,  with the upper portion becoming active before the lower part.  This suggests that  pressure  is also a  relevant parameter in the system. 
%
%the relevant parameters describing the phase diagram are strain amplitude and pressure.
%
This is reminiscent of, but different from, the case of non-Brownian suspensions~\cite{pine_chaos_2005,corte2008random}, where the phase diagram is governed by strain amplitude and density.

In marked contrast with  other theoretical and experimental absorbing state systems, for which the static absorbing configurations are disordered, our system exhibits an additional ordering transition.  This transition is manifested by crystallization, as indicated by the bond-orientational order parameter $\Psi_6$, and is characterized by singular behavior of the density of topological defects like dislocations.  To our knowledge, this is the first system where an absorbing state transition and  an ordering transition occur together, each with its own distinct order parameter.

 The experiment  consists of a monolayer of $\sim 3000$ acrylic disks, each $D = 7$ mm in diameter and 3 mm thick, confined between two glass plates and subjected to oscillatory shear with maximum strain amplitude $\gamma$.  To generate the shear, the bottom wall of the cell is moved by a  motorised translator stage; the side walls move about a fixed  pivot connected to the top wall of the shear cell. The spacing between the plates is slightly greater than the thickness of the disks, ensuring  that the disks remain in a single layer, with no buckling. The plane of the apparatus is inclined by an angle $\approx 50^{\circ}$ with respect to the vertical. This angle was chosen so as to be steeper than the angle at which sliding of the disks on the glass surface ensues, which we measured to be $\approx 27^{\circ}$. Despite this, we occasionally observe transient sticking of some disks to the plate. Lines  etched on the disks may be used as  fiducial markers to track the rotation of the disks.

 A given run consists of many shear cycles $(i) \rightarrow (ii) \rightarrow (iii) \rightarrow (iv) \rightarrow (iv) \rightarrow (i)$ as shown in Fig. \ref{fig:Fig1}. During a cycle, the system is photographed by an array of raspberry pi cameras with spatial resolution 7 pixels/mm.   We associate the configuration $(i)$ with the zero phase of the oscillatory drive. Unless explicitly stated, when we refer to  ``strobed" images, we mean photographs taken at this zero phase condition.  Before changing the strain value for a new run, the shear cell is brought horizontal to the ground and shaken, so as to erase all memory of the  previous shear cycles.

%\textcolor{red}{\it Here we should describe the system - N disks in a shear cell, with the bottom being the translation stage, and the pivot points at the top - maybe indicate them in the figure.  The disks are made of ?, the plates are made of ?, and the angle with respect to the vertical is ?.  A cycle is indicated in the figure.  Strobing?}
\begin{figure}[t]
	\centering
	\includegraphics[width=1\linewidth]{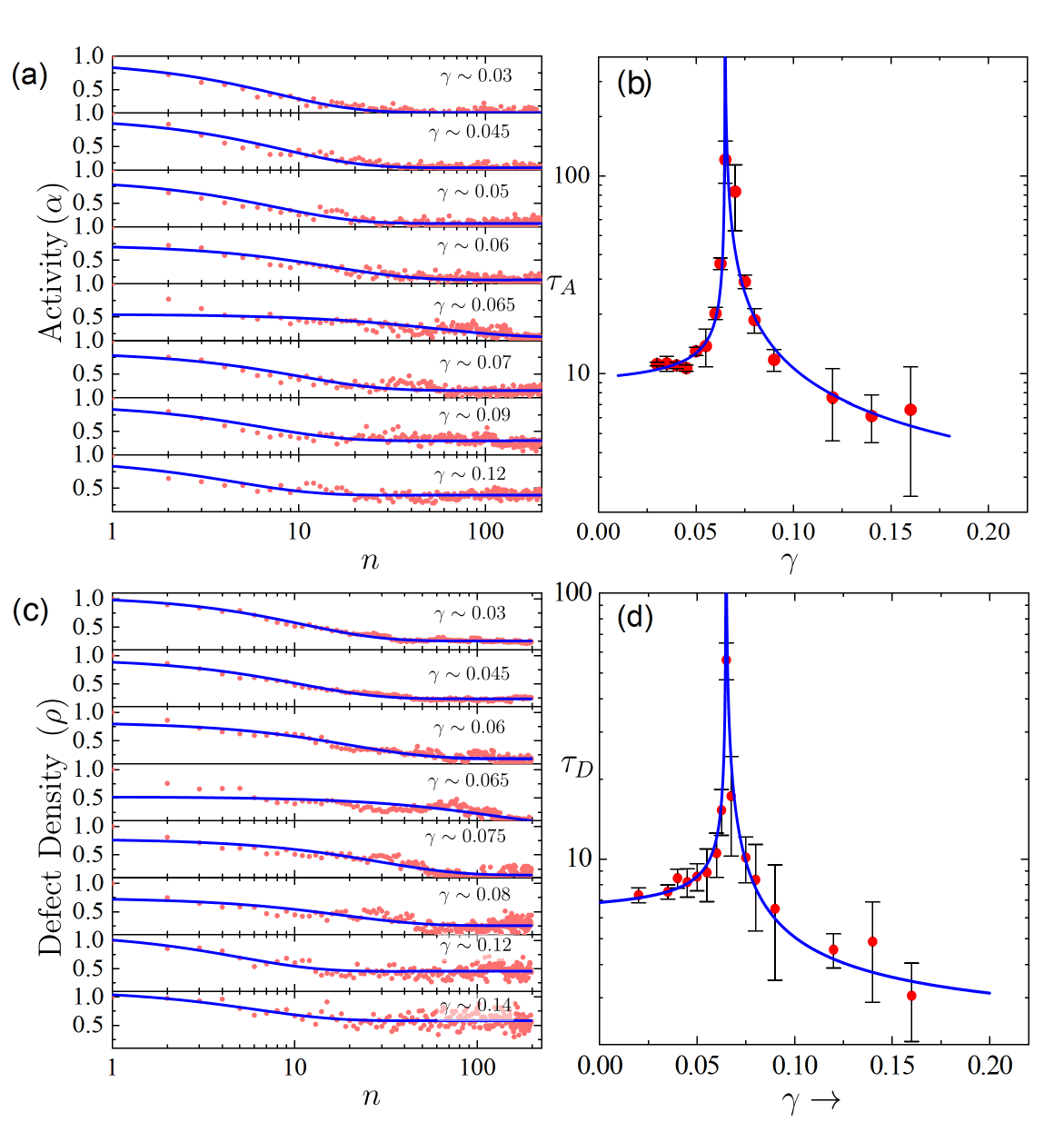}
	\caption{  (a)  Dependence of the activity $\alpha$    on the time (in number $n$ of shear cycles); panel (c)  shows the corresponding dependence of the defect density $\rho_D$.
	%, where time is measured as cycle number $n$. 
	The data in (a) and (c) are fitted with two exponential fits: $\alpha = (1-\alpha_s)\exp(-n/\tau_A)+\alpha_s$ and  $\rho_D = (1-\rho_{Ds})\exp(-n/\tau_{D})+\rho_{Ds}$, which are indicated by the blue lines in the panels. In (b) and (d) we plot the dependence of the characteristic shear cycle numbers $\tau_A$ and $\tau_D$, respectively, as a function of strain $\gamma$. In both cases, the characteristic  value of shear cycle required to reach a steady state show a divergence of the form $|\gamma - \gamma_c|^{-\nu}$ with $\gamma_c \sim 0.65$ and $\nu \sim 3/4$ for both. 
}
	\label{fig:timescale}
\end{figure}

\begin{figure}[t]
	\centering
	\includegraphics[width=1\linewidth]{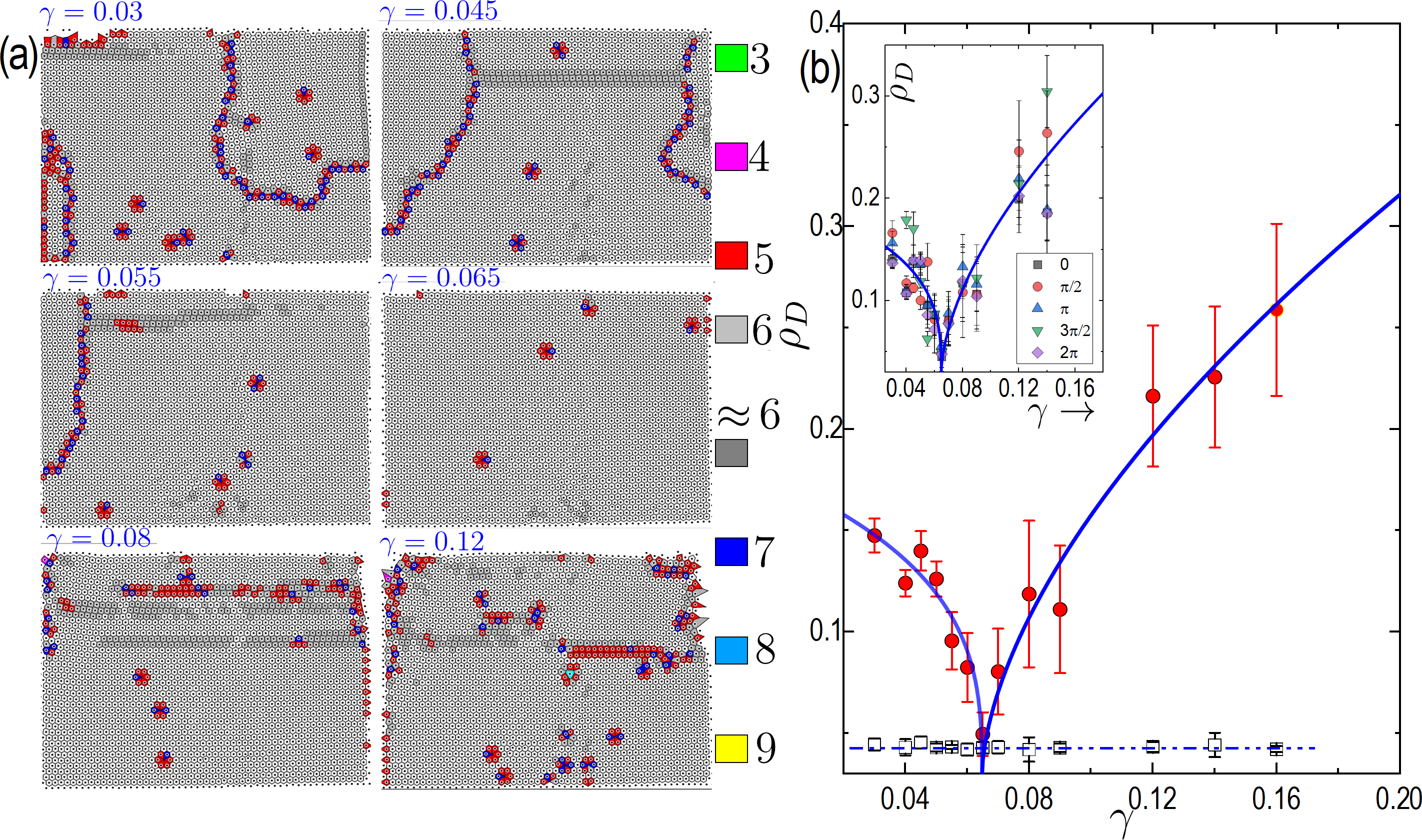}
	\caption{ (a) Characteristic snapshots of the system showing the positions of defects in the hexagonal packing, for several values of strain $\gamma$. The images are taken at the zero position in the cycle (see Fig. \ref{fig:Fig1}).  (b) Variation of the defect density $\rho_{D}$ as a function of $\gamma$. Solid red circles are for steady states obtained starting from a disordered configuration, for which a cusp-like dip in $\rho_{D}$ is seen, suggesting a second-order type transition.   The blue lines  represent the function $\rho_D \sim |\gamma - \gamma_c|^\beta$, where $\beta \sim 1/3$.  
	The hollow blue squares are for runs starting from an ordered state, for which there is no appreciable defect buildup. Inset: $\rho_{D}$ {\it vs} $\gamma$ as measured at different stages of the strain cycle. In all these cases, the initial state was disordered.
	}
	\label{fig:defects}
\end{figure}

Before focusing on the dependence on height, let us  study the active to  absorbing transition by considering measures defined on the entire system.  We consider a disk to be   active if it does not return to the same position at the end of  a cycle.
%, that is, its positions in successive strobed images are different.
Our measure of activity will be the fraction  $\alpha$ of currently active disks, which can thus be measured directly by comparing images  taken  at the beginning and at the end of a shear cycle.  For the first few shear cycles, $\alpha$ is large, since the configurations start out random, but it decreases in an exponential manner as the system self-organizes.  This is seen in Fig. \ref{fig:timescale}(a), where $\alpha$ is plotted as a function of time, measured in shear cycles, for various strains; from these data a characteristic  time   $\tau_A$  is extracted for each strain.   This is plotted in Fig. \ref{fig:timescale}(b), which shows a divergence at a critical value of the strain, $\gamma_c \approx 0.065$.  This is the first indication of a dynamical phase transition.

%Another signature of the phase transition is the cumulative mean square displacemen (CMSD), shown in Fig. \ref{MSD}.  For strains above $\gamma_c \approx 0.065$, the CSMD diverges with time, while for $\gamma < \gamma_c$, it tends to a constant.  \textcolor{red}{\it Not sure what to do with the rest of the figure - it has lots of overlap with Fig. \ref{fig:bluepic}, and I am not sure what the meaning is of $\phi_A$, since it keeps on growing with time.  We could plot the diffusion constant, I suppose.  Any thoughts?}
%In our experiments, we monitored the trajectories of each of the disks over many shear cycles.  The shear is cyclic, and we  will want to follow the motion of the disks in the different cycles.  To this end, we denote the position of the $i^{th}$ disk at time $t$ in the $n^{th}$ cycle by ${\bf r}^n_i(t) = (x^n_i(t),y^n_i(t))$, where $(x,y)$ are the in-plane coordinates of the center of a disk, and where $t$ runs over the duration of a single cycle: $0\le t\le T$.  By etching a line on each disk, we were able to follow their angular orientations.  Figure \ref{fig:NewFig2} shows the cumulative mean square displacement (CMSD) as a function of cycle number $n$:
%\begin{equation}
%    \Delta r^2_n=\sum_{i=1}^{i=N}({\bf r}^n_i(T)-{\bf r}^0_i(T))^2
%    \label{csmd}
%\end{equation}
%where the positions are sampled at the end of each cycle, $t=T$, and the sum is over all the disks, $N$ in number.

\begin{figure}[b]
	\centering
	\includegraphics[width=.9\linewidth]{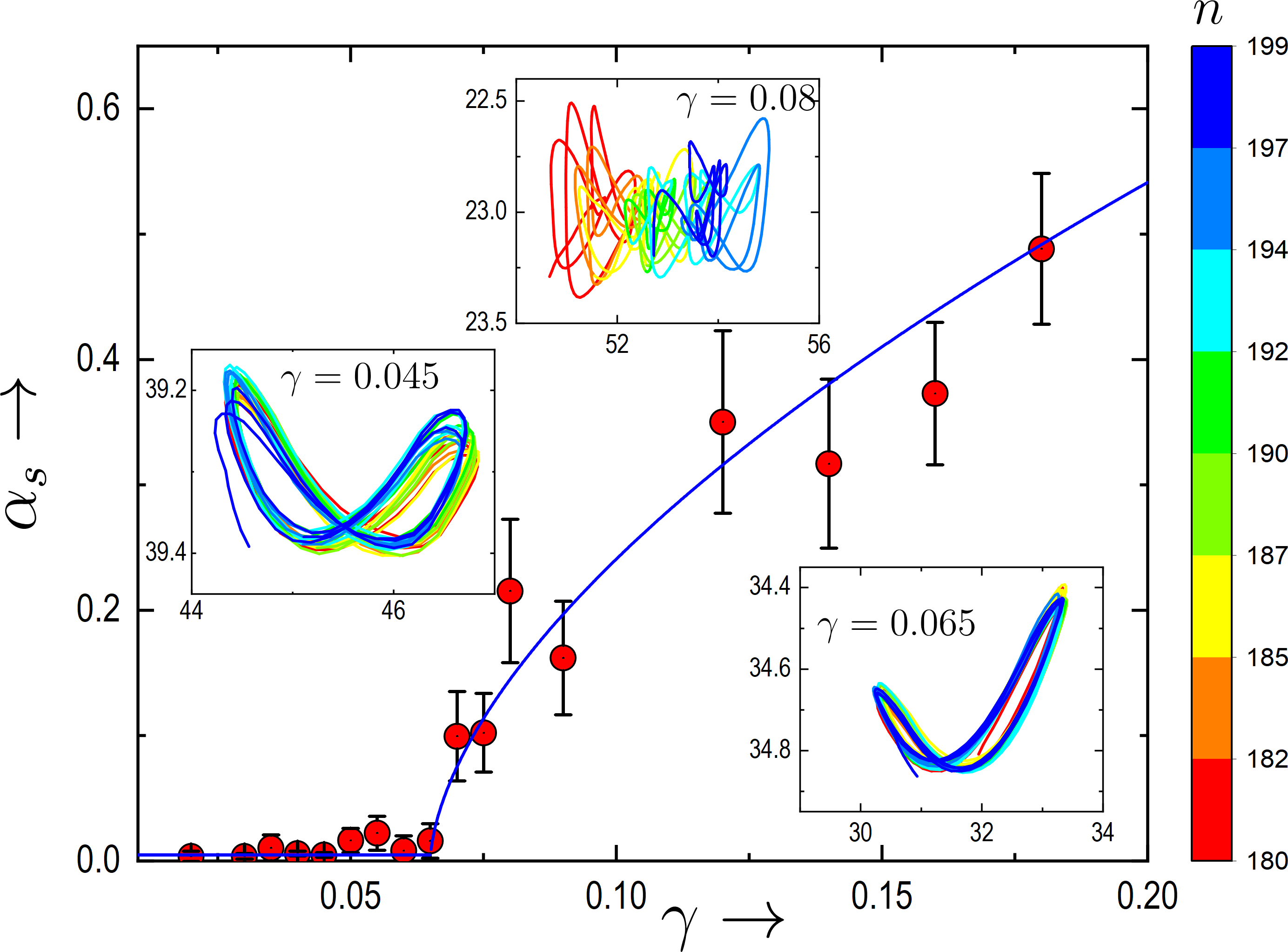}
	\caption{Saturation activity $\alpha_s$ of the system as a function of strain $\gamma$.
	%for different depths, showing the dependence of the critical strain on pressure. Note that for the smallest depth, corresponding to the pressure of a monolayer of disks, the critical gamma is $\gamma_c \approx 0.065$. 
	We find $\alpha_s = 0$ for $\gamma \le \gamma_c$, and $\alpha_s \sim |\gamma - \gamma_c|^{0.6}$ for $\gamma > \gamma_c$. Typical particle trajectories are shown in the inset for three values of $\gamma$, one above, one below, and one equal to $\gamma_c$. The colors indicate the time as measured in strain cycles.}
	\label{fig:loops}
\end{figure}

As noted earlier, in addition to the absorbing state transition, there is an ordering transition as a function of strain.  Fig. \ref{fig:defects}(a) shows Voronoi constructions of several typical configurations, at different strains, taken after the system has reached  steady state.  We identify defects as Voronoi cells with more or fewer than six edges.  The system has a tendency to order in a triangular lattice (which yield hexagonal Voroni cells), and the density $\rho_D$  of these defects tracks the crystalline order.  Moreover, the bond-orientational order parameter $\Psi_6$ is maximal at $\gamma_c$; see SI.

The dependence of $\rho_D$  on the shear cycle number for several strain values is shown in Fig. \ref{fig:timescale}(c), from which we can extract a $\gamma$-dependent characteristic time, $\tau_D$.  This is plotted as a function of strain in Fig. \ref{fig:timescale}(d), showing a divergence.  Comparing the  dependence of the activity, Fig. \ref{fig:timescale}(b) on the number of shear cycles, with that of the defect density in Fig. \ref{fig:timescale}(d), we see that the ordering and the absorbing transitions occur at the same value of $\gamma_c$.  In both cases there is similar dependence on the characteristic times for ordering, with both diverging at $\gamma_c$ as $|\gamma - \gamma_c|^{-\nu}$ with $\nu \sim 3/4$, suggesting that the two transitions are coupled. We note that the exponent we measure is very different from that of the Conserved Directed Percolation (Manna) Class, for which $\nu \approx 1.225$~\cite{Non-Equilibrium_Book}.

For $\gamma < \gamma_c$, the defects migrate together to form grain boundaries, as seen in Fig. \ref{fig:defects}(a), while more isolated defects and the clusters of defects are seen at higher $\gamma$.  We note that existence of defects is not a direct indicator of activity, which is a question of recurrent trajectories. In Fig. \ref{fig:defects}(b), the density of defects is plotted against strain, and shows a cusp-like minimum at $\gamma_c$:  $\rho_D \sim |\gamma - \gamma_c|^\beta$, with $\beta \sim 1/3$ ($\Psi_6$ also shows a cusp-like maximum; see SI.). We emphasize that these data were taken for systems which were initially disordered.  When the system begins in an ordered lattice, it maintains its order for all but the highest strains studied, as shown by the hollow blue markers in Fig. \ref{fig:defects}(b).  We note that such dependence on initial states has been seen in other absorbing state  models~\cite{hexner2017enhanced} - that is, specially chosen initial states can produce final states with different statistical characteristics than those resulting from random initial states. 

The collective behavior described above has direct counterparts in the individual particle trajectories.  For small strains, one might expect that the particle trajectories simply move back and forth in a time-reversal invariant manner.  In fact, this is not the case.  Reminiscent of the trajectories seen in computer simulations of frictional spheres in three dimensions~\cite{royer2015precisely}, we see that the trajectories form closed loops, as seen in Fig. \ref{fig:loops}.  In the absorbing phase, $\gamma < \gamma_c$, the loops are figure-8 shaped, remaining so until $\gamma_c$, while in the active phase the disks diffuse.  In the SI we present snapshots of particle trajectories over the entire system, showing a clear left-right asymmetry, as well as the differences in trajectories near to and far from the side walls.  

\begin{figure}[t]
	\centering
	\includegraphics[width=.95\linewidth]{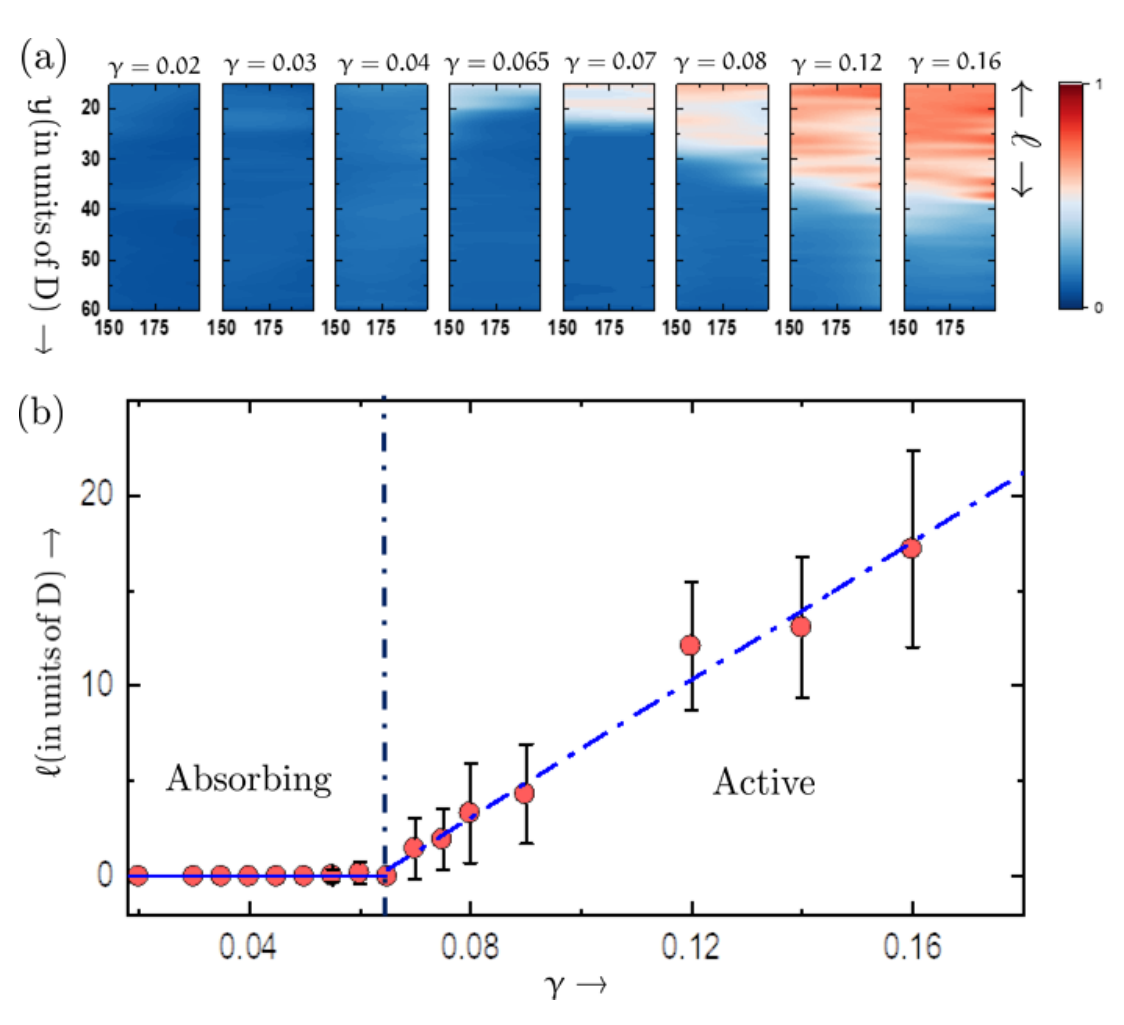}
	\caption{(a) Average activity {\it vs} depth for several values of strain during a time window spanning 50 cycles.  Blue indicates no or small activity, red indicates high activity.  For each of the panels, the activity is estimated by subtracting successive strobed images, and averaging over slices at each height (See SI). Thus, we see that the lower portion of all the systems is recurrent, and that the active region at the top increases in size as $\gamma$ increases above $\gamma_c$.  No activity is observed for $\gamma < \gamma_c$.  (b) The depth of the active region (in units of disk diameter), plotted as a function of $\gamma$. The dashed line is what might be expected, as discussed in the text.}
%	\caption{(a)Variation of $\langle \Delta I(y,n)\rangle$ for various values of $\gamma$.  Here $\langle \Delta I(y,n) \rangle = \langle |I(y,n)-I(y,n+1)| \rangle$ represents the difference in the image intensity $I(x,y)$ between two subsequent strobed images, $n$ and $n+1$. The angular brackets correspond to the averaging over the $x$ values, so that the images show average activity as a function of height.  The blue regions correspond to the periodic part of the dynamics, here after each cycle particles return to starting position. On the other hand we have the red region where the particles are active. Beyond the critical strain this active region gets initiated from the free surface and then it increases with the strain. (b) Here we plot the length $\ell$ in units of the particle diameter $D$ of the active region as a function of $\gamma$. (c) The active to non-active transition is only true for the upper half of the container, as shown in the figure. The dynamics in the cellÕs lowest part remain periodic, hence there is no transition.}
	\label{fig:bluepic}
\end{figure}

The dependence on height is seen in Fig. \ref{fig:bluepic}(a), which depicts the overlap of the configuration over time, having passed through an initial transient period during which the system organizes.  The color bar indicates the amount of overlap of successive strobed images as a function of the shear cycles, averaged over all the particles at the same height. Data is shown for 50 cycles. The bottom portions of the images are dark blue, indicating that these areas are recurrent - successive strobed images overlap almost perfectly over the course of many cycles.  These images show the essence of the system:  For small $\gamma$, essentially all the disks  return to the same positions at the zero phase point of the cycles, having traced out closed loops as seen in Fig. \ref{fig:loops}.  As $\gamma$ is increased, the system destabilizes from the top down - disks in the upper portion of the system no longer return, but those lower down do. For still larger $\gamma$, very few of the disks return, with their trajectories being diffusive.

Fig. \ref{fig:bluepic}(a) shows the average activity at depths $\ell$ as a function of the shear  cycles $n$, for different strains $\gamma$. For $\gamma < \gamma_c \approx 0.065$, we see that essentially the entire system is recurrent.  Above $\gamma_c$ there is an increase in activity beginning at the top of the system.  We suggest that for a given depth, there is some value of the strain such that disks at and above that depth will not be recurrent, but those beneath will be.  This interpretation suggests that $\gamma \approx 0.065$ is the critical shear corresponding to the weight of one layer of disks. 

A heuristic argument for the reason behind the dependence of average activity on depth can be made as follows.  Consider a typical row at depth $\ell$.    When the system is strained, the row of disks is compressed by an amount proportional to $\gamma$.  This produces a stress on the disks which itself is proportional to $\gamma$.  This stress will have an upward component on some of the disks in the row, which is resisted by the downward pressure due to the disks above it.  When the upward stress exceeds the pressure, the disks will rise.  In a coarse-grained sense, the stress at depth $\ell$ will scale as $G \gamma - \rho g \ell$, where $G$ is an effective elastic modulus, $\rho$ is the mass density of the disk configuration (roughly the disk mass times the area fraction), and $g$ is the gravitational acceleration, so that $\rho g \ell$ is the pressure on the layer.  We have neglected friction with the back plate since it is inclined more than the angle of repose.  When $G \gamma \gtrsim \rho g \ell$, some disks will displace. A fraction of these disks will not return to their initial positions, resulting in activity.  $\gamma_c$, then, would be the strain required to distort the top layer, for which $\ell \sim$ one disk diameter.

%A heuristic argument for the why the behavior depends on depth can be made as follows.  Consider a typical row at depth $\ell$.    When the system is strained, the row of disks is compressed by an amount proportional to $\gamma$.  Since the rows are not perfectly straight, this compression produces a force, which will have an upward component.  This component is resisted by the downward force due to the disks above it.  When the upward force exceeds the downward force, the disk will rise.  Since the downward force is the result of the weight of the disks above the row in question, it will result from the pressure above it, which is proportional to the pressure at that depth.  
%In a coarse-grained sense, the total force per unit length at depth $\ell$ will scale as $G \gamma - \rho g \ell$, where $G$ is an effective elastic modulus, $\rho$ is the mass density of the disk configuration (roughly the disk mass times the area fraction), and $g$ is the gravitational acceleration, so that $\rho g \ell$ is the pressure on the layer.  
%We have neglected friction with the back plate since it is inclined more than the angle of repose.  
%When the pressure can not overcome the upward force, some of the disks will rise, and may not return to their initial positions, resulting in activity.  $\gamma_c$ would be the strain required to distort the top layer, for which $\ell \sim$ one disk diameter.  

Fig. \ref{fig:bluepic}(b) shows the minimum depth where the activity is zero.  For small strain, $\gamma < \gamma_c \approx 0.065$, this is zero, since the entire system is inactive.  Above this strain, the upper portion of the configuration becomes active, with depth increasing as strain increases.  Effectively, Fig. \ref{fig:bluepic}(b) depicts the phase diagram of the system - if we interpret the depth as being pressure, then the portion above the curve is the absorbing phase where the system is recurrent, and below the curve it is active. We expect, given that contacts with the side walls are continually being broken, that little  of the weight is taken up by friction with the side walls, so the pressure should be a linear function of depth.  This picture is complimented by the images of particle trajectories shown in the SI.  Below the relevant critical height, the trajectories form (comparatively) tight orbits, giving way to increasingly fuzzy and then diffuse paths as we look higher and higher.

Last, we turn to the question of disk rotation. We have etched lines on the disks, which allows us to follow the rotation of the individual disks. The rotational motion of a disk behaves like a biased random walk, with an angle $\phi$ behaving as $\frac{d \phi}{dt} = \omega + \eta(t)$.  That is, the particle rotates with an average angular velocity $\omega$, augmented by a noise term $\eta$.  Roughly half the particles rotate in a clockwise fashion ($\omega<0$), and the rest counterclockwise ($\omega>0$). As seen in the SI, for $\gamma \le \gamma_c$, the maximum $\omega$ value found is approximately the same.  As $\gamma$ increases past $\gamma_c$, the maximum $\omega$ increases. Clockwise and counterclockwise rotators are distributed throughout the system, as is expected for frictional systems (think of interlocking gears), but a preponderance of one chirality or the other is seen at the side walls. 

We have studied a cyclically driven system of disks in two dimensions, with maximum strain acting as the control parameter.  Defining an active particle as one which does not return to its initial position at the end of a drive cycle, we find that there is a well-defined active-absorbing phase transition.  We note, however, that the transition does not appear to be of the Conserved Directed Percolation (Manna) class~\cite{Non-Equilibrium_Book}.  This may be becuase there is an additional ordering field in the problem.  In the typical CDP class, there is no spatial symmetry breaking involved, as both absorbing and active states are disordered.  Because the disks are monodisperse, they have a natural tendency to crystallize in a triangular lattice.  It is known that the addition of a second conservation law can change the nature of these states~\cite{hexner2017noise}, and we may speculate that the addition of an ordering field may play an analogous role.

%Acknowlegements:
We acknowledge support of the Department of Atomic Energy and Science \& Engineering Research Board, Government of India, under Projects 12-R\&D-TFR-5.10-0100 and CRG/2020/000507. DL thanks the Israel Science Foundation (Grant No. 1866/16) and the  U.S.-Israel Binational Science Foundation (Grant No. 2014713). PMC acknowledges support from DOE BES under grant No. DE SC0020976 for modeling and data analysis. 

\end{document}

\newcommand{\beginsupplement}{%
        \setcounter{table}{0}
        \renewcommand{\thetable}{S\arabic{table}}%
        \setcounter{figure}{0}
        \renewcommand{\thefigure}{S\arabic{figure}}%
     }
%\preprint{APS/123-QED}

\title{Supplementary Material: Coupled dynamical phase transitions in driven disk packings}% Force line breaks with \\

\author{ Akash Ghosh}
\affiliation {Dept. of Condensed Matter Physics and Materials Science, Tata Institute of Fundamental Research, Mumbai 400005, India}
\author{Jaikumar Radhakrishnan}
\affiliation{School of Technology and Computer Science,  Tata Institute of Fundamental Research,  Mumbai 400 005, India}
\author{Paul Chaikin}
\affiliation {Center for Soft Matter Research and Department of Physics, New York University, New York, NY 10003}

\author{Dov Levine}
\affiliation { Department of Physics, Technion-IIT, Haifa 32000, Israel}
\author{Shankar Ghosh  }
\affiliation {Department of Condensed Matter Physics and Materials Science, Tata Institute of Fundamental Research, Mumbai 400005, India}

\date{\today}% It is always \today, today,
%\linenumbers

\maketitle

%\newpage
\beginsupplement

\section{Experimental Details}

To construct the  trajectory of a system of $N$ disks, two tasks need to be accomplished.  First, centers of the disks must be identified in each image. This is used to generate a list of positions  $P_{t}( \mathbf {r}_1,\mathbf {r}_2, \ldots \mathbf {r}_N)$ for  every instance of  time $t$. Here, ${\bf r}_i(t) \equiv (x_i(t),y_i(t))$  is the position vector associated with the center $(x_i, y_i)$ of the $i$\textsuperscript{th} disk. Second, a   correspondence needs to  be established  between the  entries of a list $P_{t_1}$ at   time  $t_1$ to   the entries of a different list $P_{t_2}$    at   time $t_2$ .   

Procedures to   generate this linked list   rely on optimizing  a suitable  cost function  that is associated with different possible associations that can be realized across  adjacent lists \cite{chenouard2014objective}. For large number of particles this method becomes highly complex due to the growing combinatorics. To reduce  the complexity, we  used two colored disks, orange and blue. Both of these  are made of acrylic and are the same size (7 mm in diameter and 3 mm  thick),  shape and frictional properties. 

%By repeating this process across subsequent  lists, it is possible to generate the trajectories of the objects. 

\subsection*{Detecting the center of the disks}

To detect the centers of the  disks with sufficient  spatial resolution, it is necessary to use an imaging system that  has (i) a large field of view, (ii)  high spatial resolution  and (iii) low image distortion.   Single camera systems (of the non-telecentric type) that have large field of view  are also  associated  with parallax errors \cite{fan2020wide}. In the context of our experiments, it would mean that  the disks  appear as circles  only at the center of the image,  towards the edge of the image the side walls of the  disks become  visible. Parallax error of this type generates systematic errors in the detection of the  disk's center.

\subsubsection*{Image stitching }

\begin{figure}[t]
    \centering
    \includegraphics[width=1\linewidth]{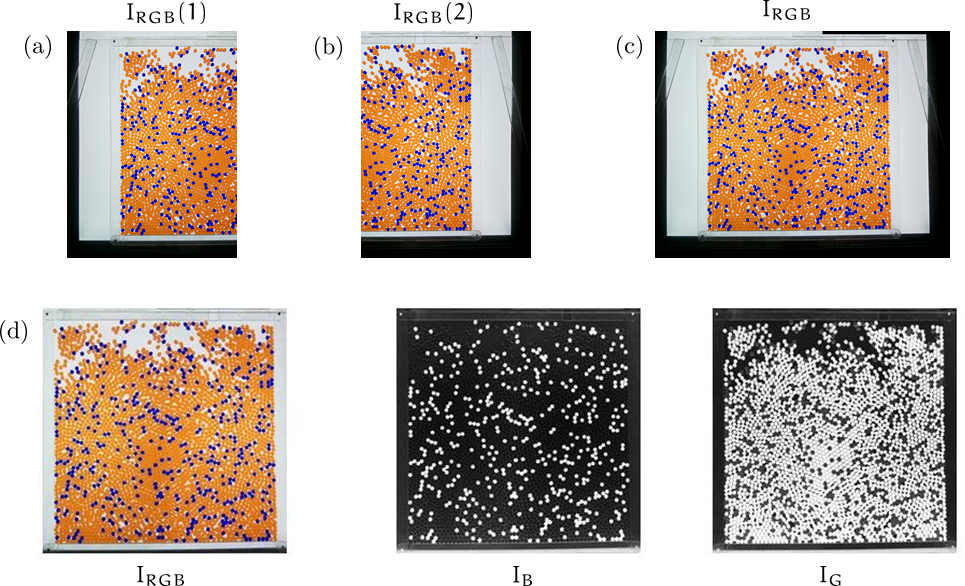}
    \caption{\textbf{Experimental Details:} The images $\mathrm{I_{RGB}(1)}$ (a) and $\mathrm{I_{RGB}(2)}$ (b)  of the  system is photographed by 2 raspberry pi cameras, each with spatial resolution 7 pixels/mm.     These images are then stitched using the  functions of the openCV stitcher  class. The stitched images are shown as $\mathrm{I_{RGB}}$  in panel (c).   Both orange and blue disks are made of acrylic and are of the same size and shape. As part of the tracking process, we separated the stitched RGB image $\mathrm{I_{RGB}}$ into three channels, the red channel $\mathrm{I_{R}}$ reveals the orange disks, whereas the blue channel $\mathrm{I_{B}}$ shows the blue disks.  We use the green channel data for normalising the intensity. Having the images partitioned into red and blue helps reduce the complexity associated with the combinatorics involved in tracking the particles.}
    \label{fig:stitch}
\end{figure}

To  achieve  the requirements of the  imaging we used two raspberry pi cameras, each with spatial resolution 7 pixels/mm.  The cameras are so positioned  that one  camera  images the left part of the system while the other camera images the right part. The central part of the system is imaged by both the cameras.  This common part is used to  stitch  the images into one composite RGB image, $I_{RGB}$. The imaging is done using the  functions of the OpenCV stitcher  class \cite{opencv_library}. Representative images captured by the two cameras and the obtained stitched image is  shown in Fig. \ref{fig:stitch}.  Please note that the stitching is not seamless  due to which an  excess vertical array of defects is often seen in the  middle  of the image. 

\subsubsection*{Image processing: } 

As part of constructing the trajectories of the disks, we  follow the  process given below
\begin{itemize}
    \item  We separate the stitched RGB image $I_{RGB}$ into three channels; the red channel $I_{R}$ reveals the orange disks, whereas the blue channel $I_{B}$ shows the blue disks.  We use the green channel $I_{G}$ data for normalising the intensity.
    
    \item  This  images $I_{R}$ and  $I_{B}$ are then  converted into their corresponding  binary images, $B_R$ and $B_B$. In these binary images the disks appear as white and the interstitial space between them  appear as black. We use  back-light illumination, this helps in  enhancing  the contrast between the white background and the colored disks. 
    
    \item  These binary images are then inverted, such that in the  inverted image  $\overline{B}_{R}$ the orange disks appear as black and the interstitial space appears as white.

    \item We then use Pythagorean  distance transform \cite{borgefors1986distance} on these inverted  binary images $\overline{B}_{R}$  and $\overline{B}_{B}$  which   creates the  distance matrices $D_{R}$ and $D_{B}$ by assigning  to each pixel in $D_{R}(i,j)$  and $D_{B}(i,j)$  the  value of the Pythagorean distance between the $(i, j)$  pixel   and the nearest white pixel in $\overline{B}_{R}$  and $\overline{B}_{B}$, respectively. Thus, the white interstitial space in $\overline{B}_{R}$   is assigned to zero in $D_{R}$ and the black pixels $\overline{B}_{R}$  are assigned to non zero values in $D_{R}$.  Center of the disk is the farthest point from  the nearest white pixels and hence have the maximum value.

    \item  We   search for local maxima in the distance matrix $D$ to find the  list of the centers  of the disks. These two lists ,one for the orange disks and the other for the blue disks are concatenated to  generate the complete position list $P( \mathbf {r}_1,\mathbf {r}_2, \ldots \mathbf {r}_N)$   of the disks. 

\end{itemize}

\section{Trajectories}

\begin{figure}[t]
	\centering
	\includegraphics[width=1\linewidth]{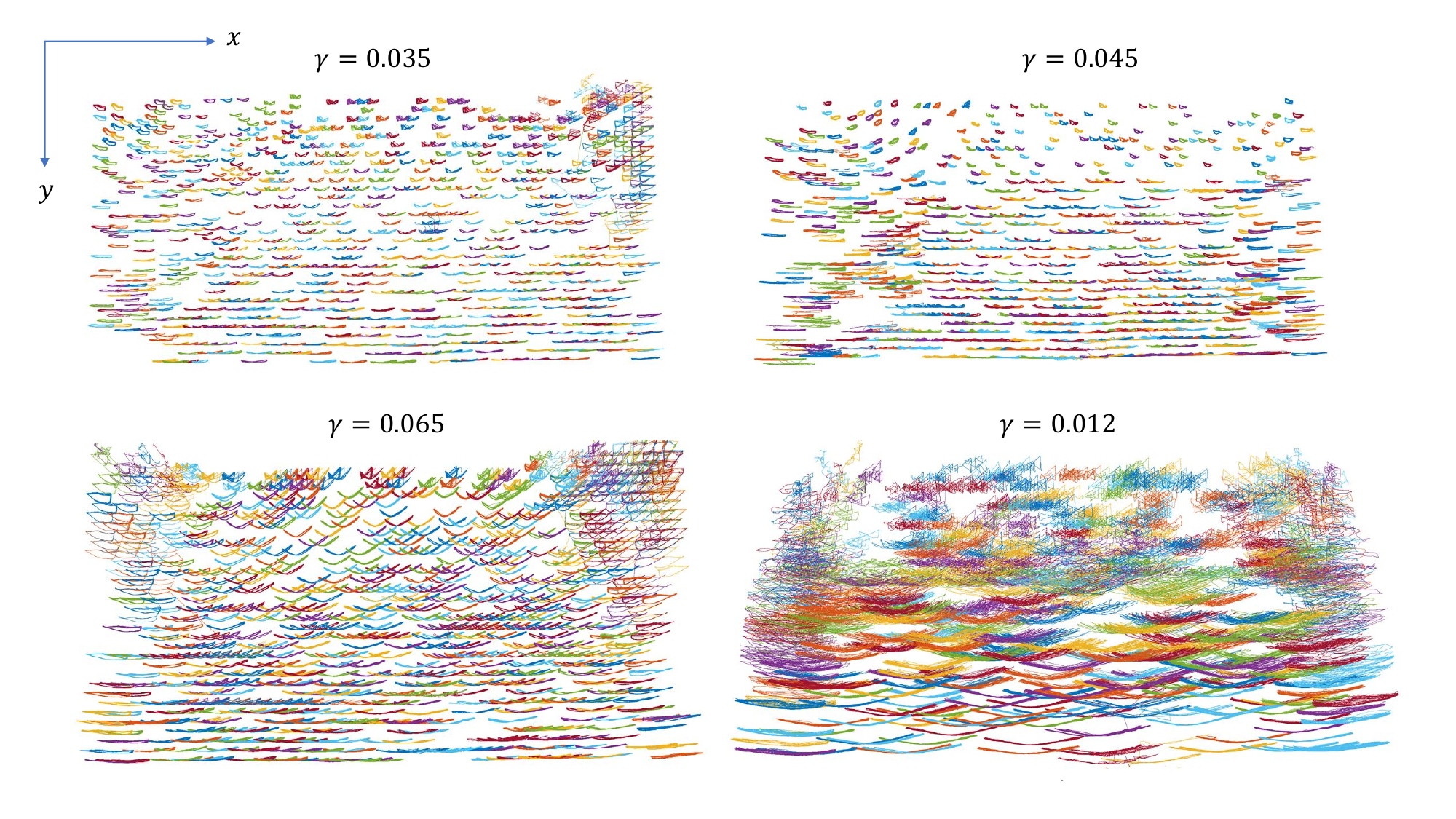}
	\caption{ \textbf{Trajectories:} The figure plots representative trajectories of  disks  for different strains $\gamma$. For $\gamma = 0.035, \gamma = 0.045$ and $\gamma = 0.065$  particle trajectories show loop like structures whereas for $\gamma = 0.12$ the trajectories are diffusive in nature.}
	\label{fig:trajectories}
\end{figure}

In our experiments, we monitored the trajectories of each of the disks over many shear cycles.  The shear is cyclic, and we want to follow motion of the disks in different cycles.  To this end, we denote the position of the $i^{th}$ disk at time $t$ in the $n^{th}$ cycle by ${\bf r}_{i,n}(t) = (x_{i,n}(t),y_{i,n}(t))$, where $(x,y)$ are the in-plane coordinates of the center of a disk, and $t$ runs over the duration of a single cycle.  By etching a line on each disk, we were able to follow their angular orientations.  Typical particle trajectories are shown in Fig. \ref{fig:trajectories} for representative  values of $\gamma$.  The colors indicate the shear cycles.  For small $\gamma$, the trajectories start out random, and gradually tighten into clearly repetitive  motion.   On the other hand, for  large $\gamma$, the  trajectories  do  not settle  down, suggesting  that  if  we  were  to  follow  them long enough they would show broadly diffusive behavior.

\begin{figure}[b]
	\centering
	\includegraphics[width=1\linewidth]{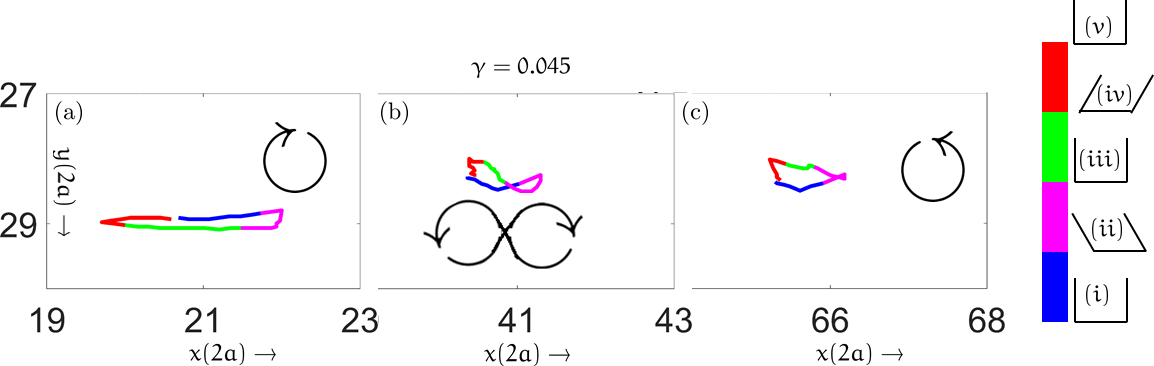}
	\caption{ \textbf{ Handedness of the trajectories} The figure plots  trajectories of a single shear cycle for   three  representative  particles, one near the left wall of the shear cell (a), one at its center (b) and the other near its right wall (c). Portion of the trajectory that corresponds to the movement of the shear cell from $(i) \rightarrow (ii)$ is  colored in blue. The next part of the trajectory $(ii) \rightarrow (iii)$  is in magenta, the third part  $(iii) \rightarrow (iv)$ is in green and the last part  $(iv) \rightarrow (i)$ that completes the shear cycle is marked red.	The particle  near the left wall of the shear cell undergoes a  clockwise motion while those near the right wall undergo an anticlockwise motion. Those that are in the center, show $\infty$-shaped loops. These particles  transit from clockwise to anticlockwise motion and vice versa during the shear cycle.
	}
	\label{fig:handedness}
\end{figure}

In Fig. \ref{fig:handedness} we explore the  handedness  of the loops associated with the  trajectories and plot paths traced out by three  representative  particles, one  near the left wall, one at the center and the other near the right wall of the shear cell. The trajectories are  plotted for a representative shear cycle.  The shear  cycle  starts from the
“zero position” $(i)$ -, moving to the rightmost position $(ii)$,
returning to the central position $(iii)$, then moving on to the
leftmost position $(iv)$ before finally coming back to the zero
position. Portion of the trajectory that corresponds to the movement of the shear cell from $(i) \rightarrow (ii)$ is  colored in blue. The next part of the trajectory $(ii) \rightarrow (iii)$  is in magenta, the third part  $(iii) \rightarrow (iv)$ is in green and the last part  $(iv) \rightarrow (i)$ that completes the shear cycle is marked red. The particle  near the left wall of the shear cell undergoes a  clockwise motion while those near the right wall undergo an anticlockwise motion. Particles in the center, show $\infty$ - shaped loops. These particles  transit from clockwise to anticlockwise motion and vice versa during the shear cycle.

\paragraph{Cumulative mean square displacement }

\begin{figure}[t]
    \centering
    \includegraphics[width=1\linewidth]{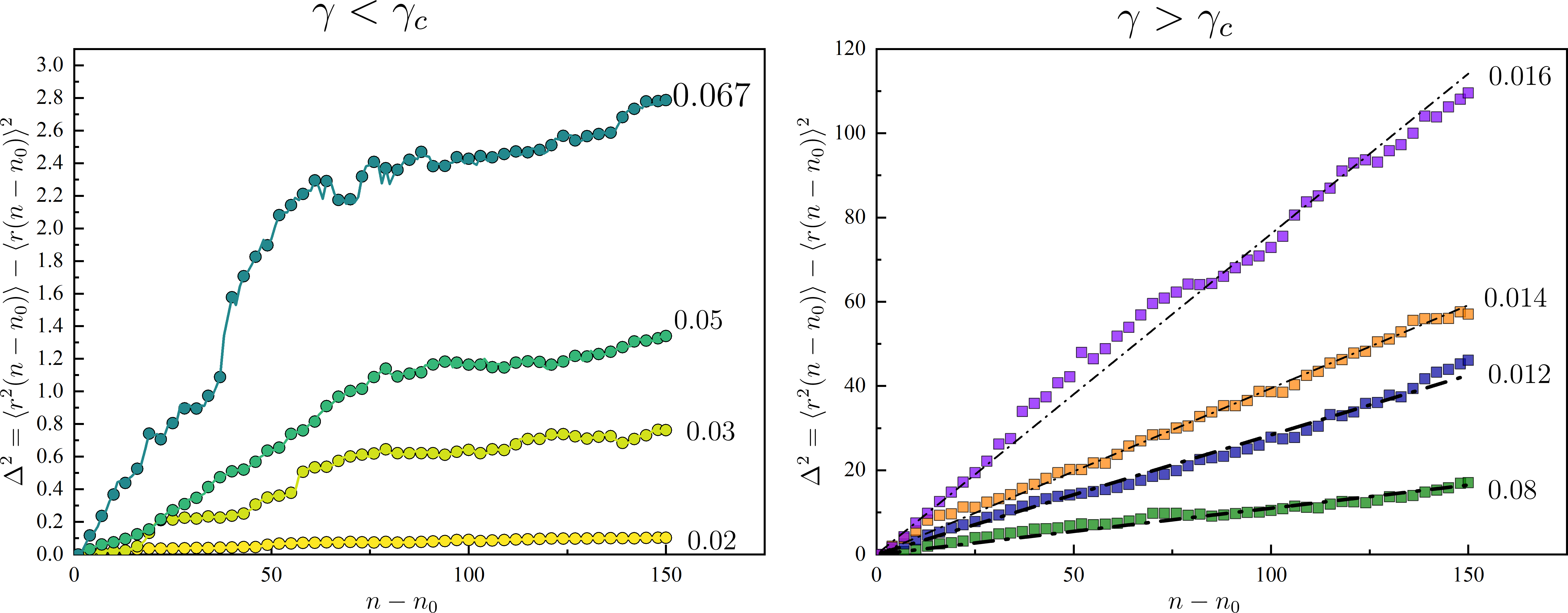}
    \caption{\textbf{Cumulative mean square displacement (CMSD):}The figure shows the cumulative mean square displacement (CMSD) as a function of cycle number $n$. The CMSD is defined as $\Delta^2=<r^2(n-n_0)>-<r(n-n_0)>^2$, where the positions are sampled at the end of each cycle n, and the sum is over all the disks.  For strains above $\gamma_c= 0.065$, the CMSD diverges with time (right panel), while for $\gamma < \gamma_c$ it tends to be a constant(left panel).}
\label{fig:CMSD}
\end{figure}

Figure \ref{fig:CMSD} shows the cumulative mean square displacement (CMSD) as a function of cycle number $n$:
\begin{equation}
    \Delta^2=<r^2(n-n_0)>-<r(n-n_0)>^2
    \label{Eqn:csmd}
\end{equation}
where the positions are sampled at the end of each cycle, $t=T$, and the sum is over all the disks, $N$ in number. For values of strain $\gamma$ that is smaller than $\gamma_c$ the  CMSD  $\Delta^2$  saturates with increasing cycle number. However,  for values of $\gamma >\gamma_c$, the dynamics of the disks becomes diffusive and  CMSD grows almost linearly with $n$. For $\gamma = 0.02, \gamma = 0.04$ and $\gamma = 0.065$  particle trajectories show loop like structures whereas for $\gamma = 0.14$ the trajectories are diffusive in nature.

% \begin{figure}
%     \centering
%     \includegraphics[width=11cm]{Pic3.png}
%     \caption{Stop-start Shearing Protocol  :  This  protocol involves moving  the  bottom wall  by (1.65 mm).   Having achieved this  displacement, the configuration of the disks are imaged. This is followed by moving the bottom plate again. The whole process of moving the bottom wall by 1.65 mm, then stopping and grabbing an image is continuously repeated. In this protocol  the  friction state of the  disks   constantly toggle from the   static to the kinetic type.}

% \end{figure}
\section{Measurement of activity}
\subsection{From the trajectories}
 The activity $\alpha$  of the system measured in the $n^{\mathrm{th}}$  shearing cycle physically means the fraction of disks  which do  not return to the position occupied by it at the beginning of the  $n^{\mathrm{th}}$ cycle.  This is measured by computing the residual displacement $\varepsilon_i^n = d_i^n$, where $d_i^n$ is the distance between the coordinates of a disk associated with its position at the beginning and end of each shear cycle. If this residual displacement $\varepsilon^n $ is greater than a  tolerance value (chosen to be 0.25 times the diameter of the disk), we consider that the particle has not returned to its earlier configuration and hence is an active particle.  The activity $\alpha$ is given by the number of active disks normalized by the total number of disks in the system.
 
 Spatial regions containing the active particles are referred to as active regions.   To estimate the  spatial structure of the  active region in the system we performed two  analysis. The first method consisted of overlaying the position of the  detected particles in the strobed condition  - for which the phase of the shear cycle is zero. This overlay is done for the last few  shear cycles. In the second method  we analysed the difference matrix  obtained by subtracting  two strobed images from subsequent shear cycles. 
 
 \begin{figure}[b]
    \centering
    \includegraphics[width=.95\linewidth]{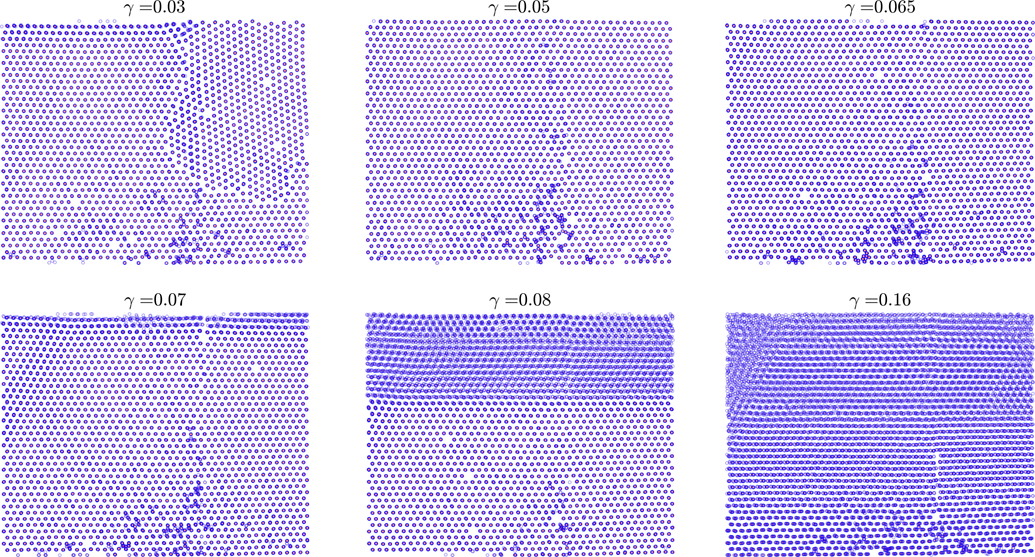}
    \caption{Representative scatter plots of the points in the set  $\mathcal{P}=\bigcup_{n=190}^{200}P_{i}$ for representative values of the strain $\gamma$.  Here  $P_n((x_1,y_1),(x_2, y_2)), \ldots (x_N, y_N))$ is  the  set of coordinates of the disks in the strobed condition in the n\textsuperscript{th} strain cycle. Please note that the vertical array of defects
that one sees at the middle is an artifact of image stitching.}
    \label{fig:strobed_overlay}
\end{figure}
\subsection{From strobed positions}
Let $P_n((x_1,y_1),(x_2, y_2)), \ldots (x_N, y_N))$ be the  set of  coordinates of the disks in the strobed condition ( phase of the stain cycle is zero)  for the n\textsuperscript{th} cycle.  We generate a larger set of points  $\mathcal{P}$ by taking the union of these sets from  different strain cycles $\mathcal{P}=\bigcup_{n=m}^{m'}P_{i}$, here $m$ and $m'$ represent the two ends of the interval $[m, m+1, \ldots , m']$ of strain cycles  which is considered in constructing the set $\mathcal{P}$.  If the system returns to its  previous position at the end of each strain cycle, then the cardinality of the set  $\mathcal{P}$ equals that of the individual position sets $P_n$. However, if there is activity in the system,  the cardinality  $|\mathcal{P}|$  increases with the  size of the interval, $m'-m$. A visual inspection of  scatter plot of all the points in the set $\mathcal{P}$  is highly instructive in uncovering the spatial structure associated with active regions. Representative scatter plots of the points in $\mathcal{P}$ for different values of strain $\gamma$ is plotted in Fig. \ref{fig:strobed_overlay}.  The scatter plot is localised in non-active parts, whereas it is smeared in the active zone.   For small values of $\gamma$ , the entire system  is  non-active. However, beyond a critical strain, the top part of the  system becomes active. This is the regions  where the  scatter plots  get smudged.  The extent of smudging grows as the amount of strain increases.

\subsection{From the strobed images}
\begin{figure}[t]
    \centering
    \includegraphics[width=.95\linewidth]{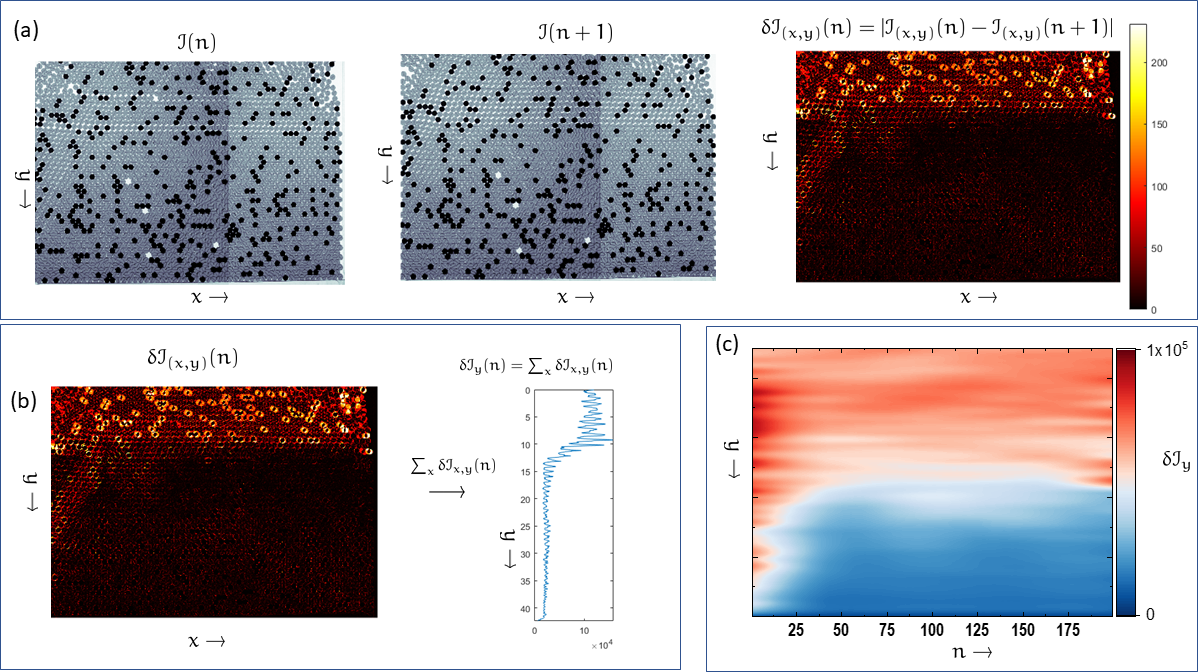}
    \caption{(a) The left and the center panels  show  strobed images for two consecutive strain cycles (n=150 and n=151 for $\gamma=0.08$). The right panel shows the   difference between these two images,  $\delta I_{(x,y)}(n)=|I_{(x,y)}(n)-I_{(x,y)}(n+1)|$. The colorbar  represents the magnitude of this difference .  (b) The  difference matrix $\delta I_{(x,y)}(n)$ is then summed over the  columns  to give a column vector $ \delta I_{y}(n) = \sum_{x}\delta I_{x,y}(n)$. This is  done for all  values of  n, as a  result we   obtain n column vectors $ \delta I_{y}(n) $. (c) These column vectors are then concatenated to generate the matrix   $\delta\mathcal{I}= (
  \delta I_{y}(1)|
  \delta I_{y}(2)|\ldots|
  \delta I_{y}(n)|
  \delta I_{y}(n+1)\ldots) $. }
    \label{fig:image_activity}
\end{figure}
   From the set of the images that were taken during the experiment we formed a  list of  strobed images $I(n)$, one taken at a predefined phase from each  shear cycle. Here, $n$ is the shear cycles number. An estimate of the activity in the system can be evaluated  by taking the difference between two consecutive  strobed images $\delta I_{(x,y)}(n)=|I_{(x,y)}(n)-I_{(x,y)}(n+1)|$ (see Fig.\ref{fig:image_activity}(a)).  Elements of the matrix $\delta I_{(x,y)}(n)$ whose value is   beyond a threshold value is  considered to be  active.  We find that these active regions are mostly  localised near the free surface, while the region close to the bottom wall is non-active.
 
 To study  the growth and organisation  of the active region in the system  as a function  of the  shear cycle  number, we first convert  the matrix  $\delta I_{(x,y)}(n)$ to a column vector $\delta I_{y}(n)$  by  summing over the columns. 
  \[
 \delta I_{y}(n) = \sum_{x}\delta I_{x,y}(n).
 \]
 The intensity profile  $\delta I_{y}(n) $,  represents the  variation of the average activity as a function of height (see Fig.\ref{fig:image_activity}(b)). Having obtained the average activity profile as a function of height for a given  value of n, we  concatenate    $\delta I_{y}(n)$ arrays vertically  for increasing values of $n$ and 
 
 \[  \delta\mathcal{I}= (
  \delta I_{y}(1)|
  \delta I_{y}(2)|\ldots|
  \delta I_{y}(n)|
  \delta I_{y}(n+1)\ldots) .
 \]
This matrix is plotted in   Fig.\ref{fig:image_activity}(c). The blue regions correspond to the periodic part of the dynamics (non-active region, here after each cycle particles return to starting position. On the other hand we have the red region where the particles are active.

\section{Quantifying  order}
We used both voronoi construction and bond orientation order parameter $\psi_6$  to estimate the  amount of disorder in the system.

\paragraph{Voronoi Construction: }
 The voronoi edges are perpendicular bisectors to adjacent sites and hence the voronoi cell is a convex polygon. Since the  number of edges $n_{E}$ of a polygon are  same as the number of its adjacent sites, we associate $n_{E}$ with the  coordination number of the point that is contained inside the cell. Here each point corresponds to the center of a disk.  In two dimension, a crystal lattice is associated with six nearest neighbours. Therefore, the voronoi cells with edges not equal to six are considered as defects. However, there are a lot of irregular hexagons associated with configurations generated by slipping of one particle layer over another. Although these hexagons have six edges, they don't correspond to a crystal and we count them  as defects. The irregularities of the hexagons are defined by the ratio of length of the longest edge to length of the shortest edge of the voronoi cell. If the ratio is more than 3.5, the corresponding hexagon is considered as a defect.

\begin{figure}[t]
    \centering
    \includegraphics[width=1\linewidth]{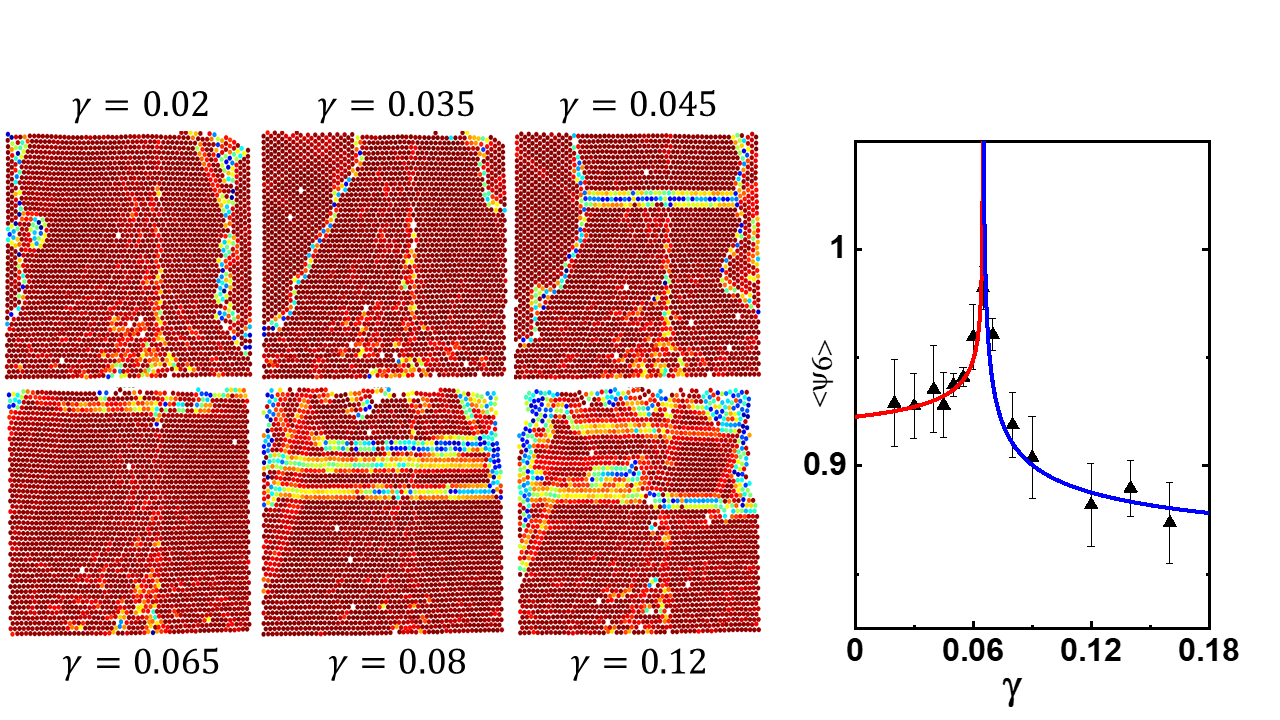}
    \caption{ (a) : The  particles are color coded in the figures. The colors  represent  the single-particle bond orientation parameter for the i\textsuperscript{th} particle and its $N_i$ nearest neighbors, $q_6(i)=\frac{1}{N_i}\sum_{j=1}^{N_i} e^{i6\theta_{ij}}$. Where $\theta_{ij}$ is the angle between the center-to-center vector from particles i to j about a fixed arbitrary  axis.  A particle colored in red is in the perfect crystalline environment $q_6(i)=1$.   As the color changes from red to blue the degree of crystallinity decreases. (b) : The degree of crystallinity for an entire configuration was quantified  via the bond orientation order parameter $\psi_6$ in the following way: $\psi_6=\langle |q_6(i)|^2\rangle$.  The  figure shows the variation of the parameter $\psi_6$ with $\gamma$. The value of $\psi_6$ is maximum as $\gamma$=0.065 indicating that the system is most crystalline for that parameter of strain. The blue lines  represent the function $\Psi_6 \sim |\gamma - \gamma_c|^\beta$, where $\beta \sim 1/3$. Please note that the vertical array of  defects  that one sees in the middle is an artifact of image stitching.}.
    \label{fig:psi_6}
\end{figure}

\paragraph{Bond orientation parameter:}  The extent of local order of the  i\textsuperscript{th} particle and its $N_i$ nearest neighbors is estimated from the single particle bond orientation parameter, $q_6(i)=\frac{1}{N_i}\sum_{j=1}^{N_i} e^{i6\theta_{ij}}$. Where $\theta_{ij}$ is the angle between the center-to-center vector from particle i to j around a fixed arbitrary  axis. In the perfect crystalline environment $q_6(i)=1$.   The degree of crystallinity decreases  with increasing values of  $\gamma$. An estimate of the degree of crystallinity for an entire configuration was quantified  via the bond orientation order parameter $\psi_6$ in the following way: $\psi_6=\langle |q_6(i)|^2\rangle$.  The  Fig.\ref{fig:psi_6} shows the variation of the parameter $\psi_6$ with $\gamma$. The value of $\psi_6$ is maximum as $\gamma$=0.065 indicating that the system is most crystalline for that parameter of strain.

\begin{figure}[h!]
	\centering
	\includegraphics[width=1\linewidth]{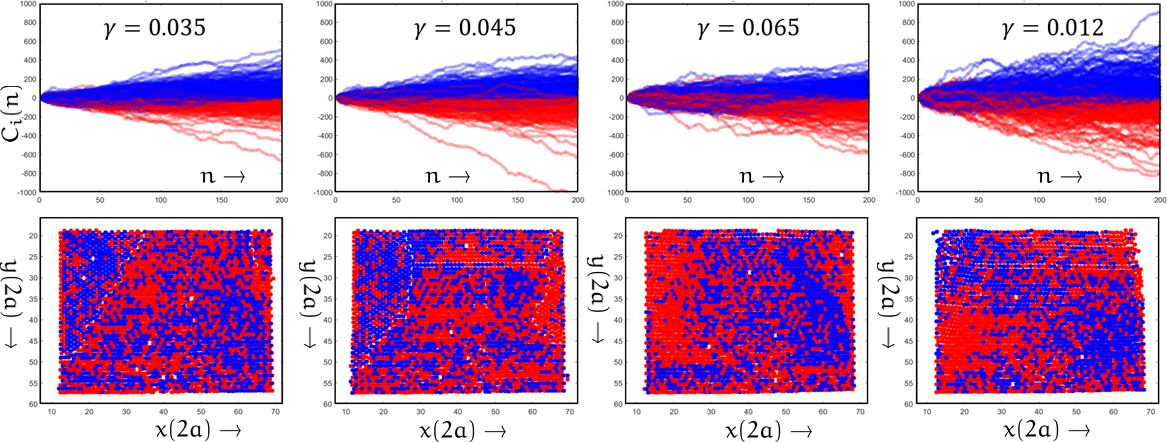}
	\caption{ \textbf{Top panel :} To describe the dynamics of the angular variable associated with the disks we keep track of the cumulative angular displacement $\vartheta$ of the disks. Here we plot the cumulative rotation of each particle with the number of shear cycles $n$ for representative values of $\gamma$. For large values of $ n $ most  disks were  found to either rotate persistently in the clockwise or counter-clockwise directions.  For disks that pick up a positive cumulative angle $C_i(n)=\Sigma_{m=1}^{m=n} \vartheta_{(i,m)}$, its trajectory, $ \Sigma_m \varphi_{{i,m}} $ is coloured red. However, if it  picks up a negative cumulative angle the  associated trajectory is marked blue. The angles are measured for the $\vartheta=0$ position (central position) of the shearing cycle.    With increasing value of $n$, the red  and the blue trajectories progressively move away from each other. \textbf{Bottom panel :} The figures show the spatial organization of the clock-wise and anti-clockwise rotating particles. The blue filled circles identify the clockwise rotating disks while the red filled circles identify the disks     that rotate in an anti-clockwise direction.  The  positions of the circles  correspond to the  the configuration attained by the disks at the end of the 200\textsuperscript{th} cycle.  }
	\label{fig:rotation}
\end{figure}

\section{Rotation of the  discs}
% \begin{figure}
%     \centering
%     \includegraphics[width=12cm]{Pic9.png}
%     \caption{Lyapunov exponent is defined as,\\
%     $\lambda_{max} = \frac{1}{t} \sum_{i=1}^{N} \sum_{j=1}^{M} \frac{1}{MN}\ln|\frac{X_i(t) - X_j(t)}{X_i(t_0) - X_j(t_0)}|$
% ,where M = number of nearest neighbors, N = Total number of particles.
% The exponent is zero for $\gamma < \gamma_c$. Beyond $\gamma_c$, $\lambda_{max}$ increases. This tells us that below $\gamma_c$ the trajectories are periodic.
% The blue curve is fitted with the data as
% $\lambda'_{max}=0.24(\gamma - \gamma_c)^{0.64}$.
% There are three representative plots of the trajectories of the particles corresponding to $\gamma = 0.03, \gamma = 0.065$ and $\gamma = 0.08$. The data is shown for last 20 cycles. The trajectories below $\gamma_c$ show loopy behaviour whereas above the transition point some of the trajectories become diffusive.
% }
% \end{figure}

We now describe the angular variable associated with the disks. To  do so we keep track of the cumulative angular displacement of the disks. For large values of $ n $ if the disks pick up a positive cumulative angle, its trajectory, $ C_i(n)=\Sigma_n \varphi^i_n $ is coloured red. However, if it  picks up a negative cumulative angle the  associated trajectory is marked blue. The angles are measured for the $\vartheta=0$ position (central position) of the shearing cycle.   In  the top panel of Fig.\ref{fig:rotation}  we plot the cumulative rotation of each particle with the number of shear cycles $n$ for different values of strain amplitude. With increasing value of $n$, the red  and the blue trajectories progressively move away from each other. The shear cycle creates a partitioning of the disks into the clockwise (blue) and the anticlockwise (red) rotating disks (see the bottom panel of  Fig.\ref{fig:rotation} ).

The persistence of rotation of the disks comes from the following argument that holds true for all dry granular systems that are subjected to the form of shearing  used in this experiment.   As the two side walls $ad$ and $bc$ synchronously move so as to increase the shearing angle,  one of the side walls become the pusher and moves  the disks while the other wall (confining wall) occasionally moves ahead of the disks creating a gap. This void is  filled  stochastically  by  disks  toppling into it. The disks near the pushing wall rotate due to the torque exerted on it by the boundaries while those near the confining wall rotate due to its toppling motion. Rotation is transmitted to the neighbouring disks in an opposite sense, i.e., a clockwise rotating disk imparts anti-close wise rotation to the neighboring disks which in turn pass on a clockwise rotation to the disks in contact, this leads to geometric frustration which  prevents transmission of rotational motion into the bulk.
In the part of the shear cycle when the pushing side is the right side wall, the disks adjacent to it pick up an clockwise motion.  Disks near the confining side wall rotate but for an entirely different reason.  Since the  origin of rotation is different for the disks near the pushing and pulling walls, the rotation picked up in one part of the  cycle is not undone by  the other part. This leads to a ratcheting  motion of the disks: on an average, the disks near the left side wall are clockwise rotating  while those near the right side wall rotate in a clockwise manner and as a result, the net rotation grows with $ n $.

%\bibliography{suppl}
